\documentclass[12pt,a4paper]{article}
\usepackage{a4wide}
\usepackage{latexsym}
\usepackage{epsf}
\usepackage{amssymb}
\begin{document}
\def\be{\begin{equation}}
\def\ee{\end{equation}}
\def\bea{\begin{eqnarray}}
\def\eea{\end{eqnarray}}

\def\pd{\partial}
\def\a{\alpha}
\def\b{\beta}
\def\g{\gamma}
\def\d{\delta}
\def\m{\mu}
\def\n{\nu}
\def\t{\tau}
\def\l{\lambda}

\def\s{\sigma}
\def\e{\epsilon}
\def\scri{\mathcal{J}}
\def\cM{\mathcal{M}}
\def\tcM{\tilde{\mathcal{M}}}
\def\RR{\mathbb{R}}
%%%%%%%%%%%%%%%%%%%%%%%%%%%%%%%%%%%%%%%%%%%%%%%%%%%%%%%%%%%%%%%%%%%%%%

\hyphenation{re-pa-ra-me-tri-za-tion}
\hyphenation{trans-for-ma-tions}

%%%%%%%%%%%%%%%%%%%%%%%%%%%%%%%%%%%%%%%%%%%%%%%%%%%%%%%%%%%%%%%%%%%%%%

\begin{flushright}
IFT-UAM/CSIC-99-46\\
hep-th/9911202\\
\end{flushright}

\vspace{1cm}

\begin{center}

{\bf\large The Master Gauge String  }

\vspace{.5cm}
 
{\bf Enrique \'Alvarez \ddag}
\footnote{E-mail:enrique.alvarez@cern.ch}
and {\bf C\'esar G\'omez\dag  }
\footnote{E-mail:cesar.gomez@uam.es}
\vspace{.3cm}

\vskip 0.4cm

{\it \ddag Theory Division, CERN,1211 Geneva 23, Switzerland,\\
 \dag\ddag Instituto de F\'{\i}sica Te\'orica, C-XVI,
\footnote{Unidad de Investigaci\'on Asociada
  al Centro de F\'{\i}sica Miguel Catal\'an (C.S.I.C.)}
and  Departamento de F\'{\i}sica Te\'orica, C-XI,\\
  Universidad Aut\'onoma de Madrid 
  E-28049-Madrid, Spain }

\vskip 0.2cm

\vskip 1cm

%%%%%%%%%%%%%%%%%%%%%%%%%%%%%%%%%%%%%%%%%%%%%%%%%%%%%%%%%%%%%%%%%%%%%%

{\bf Abstract}
 A string background, which is in some precise sense {\em universal}
(i.e., incorporating all orders in the Feynman diagram expansion),
is proposed to represent pure gauge theories.

\end{center}
%%%%%%%%%%%%%%%%%%%%%%%%%%%%%%%%%%%%%%%%%%%%%%%%%%%%%%%%%%%%%%%%%%%%%%

\begin{quote}

\end{quote}

%%%%%%%%%%%%%%%%%%%%%%%%%%%%%%%%%%%%%%%%%%%%%%%%%%%%%%%%%%%%%%%%%%%%%%

\newpage
%%%%%%%%%%%%%%%%%%%%%%%%%%%%%%%%%%%%%%%%%%%%%%%%%%%%%%%%%%%%%%%%%%%%%%

\setcounter{page}{1}
\setcounter{footnote}{1}
%\renewcommand{\theequation}{\thesection.\arabic{equation}}
%\tableofcontents
\newpage

\vspace{1cm}
%%%%%%%%%%%%%%%%%%%%%%%%%%%%%%%%%%%%%%%%%%%%%%%%%%%%%%%%%%%%%%%%%%
%\section*{1}
%%%%%%%%%%%%%%%%%%%%%%%%%%%%%%%%%%%%%%%%%%%%%%%%%%%%%%%%%%%%%%%%%%%%
Vacuum configurations for the closed bosonic string are determined
by the sigma model beta function equations {\cite{callan}},
{\cite{curci}}
. These are the equations that ensure
world sheet conformal invariance. At tree level in string perturbation theory
these equations, for vanishing tachyon expectation value, are given by:
\be\label{metric}
R_{{\mu}{\nu}}+{\nabla}_{\mu}{\nabla}_{\nu}{\Phi}=0
\ee
\be\label{escalar}
({\nabla}{\Phi})^{2}- {\nabla}^{2}{\Phi} = C
\ee
With ${\Phi}$ the dilaton field and the constant $C$ being determined
by the target space-time dimension as follows:
\be
C= \frac{26-D}{3}
\ee
The renormalization group approach to the string representation of
gauge theories
{\cite{alvarezgomez1}}, {\cite{alvarezgomez}},{\cite{akhmedov}} is based on:
\par
i) Identifying the string coupling constant with the Yang Mills
coupling:
\be\label{g}
g=e^{\frac{\Phi}{2}}
\ee
\par
ii) Identifying one space-time coordinate with the renormalization
group variable ${\mu}$, and
\par
iii) Looking for vacuum solution to equations ({\ref{metric}})({\ref{escalar}})
and to the renormalization group equation:
\be\label{beta}
{\mu}\frac{dg}{d{\mu}} = {\beta}(g({\mu}))
\ee
for ${\beta}$ the Yang Mills beta function.
\par
In this letter we will present a critical ,i.e $C=0$, universal
solution to equations ({\ref{metric}}),({\ref{escalar}}),({\ref{beta}}) for 
${\beta}(g)$ a generic formal power series of $g$ of the type:
\be\label{beta}
{\beta}(g) = \sum_{i}b_{i}g^{i}
\ee
In terms of the dilaton field ${\Phi({\mu})}$ defined by 
equations ({\ref{g}}) and ({\ref{beta}}) the universal metric is given by:
\be\label{universal}
{ds}^{2}= e^{{\Phi}}{d\vec{x}}_d^{2} + (d e^{{\Phi}})^{2} + 
{d\vec{y}}_D^{2}
\ee
Before going into a brief discussion on the physical meaning
of this string metric let us prove that metric ({\ref{universal}})
with $D=21$ and a dilaton
field given be ${\Phi({\mu})}$ is in fact solution to equations 
({\ref{metric}})
and ({\ref{escalar}}) for $C=0$.
\par
The proof of our previous statement follows directly from the expression
for the Ricci tensor for the universal metric
\bea
R_{\m\n}&=& -\frac{1}{4} \frac{2 a'^2 b - a'b' a + 2 a''ba}{b^2 a }\eta_{\m\n}
\nonumber\\
R_{44}&=&-\frac{-a'^2 b + 2 a'' b a - a' b' a}{a^2 b}
\eea
(where $a\equiv e^{ 2 \Phi}$ and $b\equiv e^{ 2 \Phi}\Phi'^2$).
The most illustrative condition comes, however, from the beta 
function of the dilaton,$\b_{\Phi}=
 (\nabla \Phi)^2- {\nabla}^2 \Phi $, which can be easily shown to be equal to 
$e^{-2\Phi}(2- d/2)$).
\par
It is important to notice that the metric ({\ref{universal}}) is solution
of the sigma model beta function equations only if $D=21$ with the other
five coordinates having the meaning of the four dimensional space-time 
coordinates and the renormalization group scale. Moreover the
universal metric is solution independently of the sign of the beta
function i.e both for asymptotically or infrared
free theories. The difference between them is related to
the meaning of the singularity of the metric 
({\ref{universal}})
at  $e^{{\Phi}} = 0$.
\par
At the level of the metric the {\em $S$-duality} transformation
can be implemented as:

\be\label{S}
{\Phi} \to -{\Phi}
\ee
Hence we can define the $S$-dual metric by performing the transformation
({\ref{S}}) on the universal metric. From a physical point of view this
$S$ dual metric could correspond to the same gauge theory but
described in terms of 't Hooft's loop variables.
\par
It would be convenient for future discussion to define the universal metric
as the five dimensional piece of the metric ({\ref{universal}}) and  
to consider this metric as solution to the sigma model
beta function equations with $C=0$. 
An important property of this five dimensional metric
is that it is conformally flat. Introducing new coordinates $x'=2x$ and 
$z^{2}=e^{{\Phi}}$ the five dimensional metric becomes:
\be\label{conf}
{ds}^{2}= 4z^{2} ({d\vec{x'}}_d^{2} + dz^{2})
\ee
with a dilaton field
\be
\Phi(z) = -lnz^{2}
\ee
Different input Yang Mills beta functions correspond to changes
of variables
\be
z \to z({\mu})
\ee
with ${\mu}$ the renormalization group coordinate.
\par
It is worth noticing the difference between metric ({\ref{conf}}) and the AdS
metric in five dimensions
\be\label{ads}
{ds}^{2}= z^{-2} ({d\vec{x}}_d^{2} + dz^{2})
\ee
These two conformally flat metrics are special in the following sense. In
the AdS case the Ricci tensor which is determined in terms of the
conformal factor becomes exactly equal to a cosmological constant term. In the
case ({\ref{conf}}) the Ricci tensor becomes exactly equal to
${\nabla}_{\mu}{\nabla}_{\nu}{\Phi}$ for ${\Phi}$ the logarithm of the 
conformal
factor in ({\ref{conf}}). We feel these two metrics are the natural
candidates to describe gauge theories with vanishing 
{\cite{malda}},{\cite{witten}},{\cite{poly}} and non vanishing
beta functions respectively\footnote{ An important difference between AdS 
metric and metric ({\ref{conf}}) is the existence of a singularity for
the metric ({\ref{conf}}) in contrast with the horizon in the AdS case.} .
Notice that in our framework ,based on the closed bosonic string,
the case of vanishing beta function is a bit special. Namely in this case
the only solution is flat Minkowski in $26$ critical dimensions. In other
words for $C=0$ and constant dilaton the equations ({\ref{metric}}) and
{\ref{escalar}} do not determine in any natural way the dimension
of the space-time in contrast to the case where the beta function 
does not vanish, where four dimensional space-time is singled out. Moreover
we should expect, from physical grounds
based in Wilson loop computations {\cite{malda2}}
,{\cite{witten2}}, that the conformally invariant situation,
, corresponding to  vanishing
beta function will determine string metrics
that are invariant under appropiated conformal rescalings of the coordinates,
which is not the case for flat Minkowski metric ( as opposed to $AdS_5$,
which enjoys the conformal group, $SO(2,4)$, as its isometry group).
In this sense please notice that
the universal metric (\ref{conf}) is not invariant under rescalings
of the coordinates, which is only natural owing to our
  non trivial dilaton background
\footnote{ The problem with the string representation of
pure gauge theories with vanbishing beta function i.e with
pure Maxwell theory of free photons, could be probably be 
related to a similar problem
appearing in the AdS approach,  concerning the physical
interpretation of the near horizon geometry of just one D-brane (cf.
(\cite{alvarezgomez0}). In heuristic
physical terms the problem has its origin in trying to represent in terms
of gravity a purely linear theory {\cite{witten}} }.
\par
The predictive power of the renormalization group approach to the 
string description of pure gauge theories should be contained in those 
dynamical
aspects of the gauge theory that we can describe in terms of what it is
properly speaking the string output of our construction, namely the string
space-time metric. The first prediction in this sense is related with
the so called Zamolodchikov's $c$-theorem. It was proved in reference
\cite{zamolodchikov}, for two dimensional quantum field theories, the existence
of a $c$-function ,roughly counting the number of degrees of freedom, that
monotonically decreases along the renormalization group flow. In reference
{\cite{alvarezgomez1}} ( see also
{\cite{bala}},{\cite{warner}},{\cite{malda3}},{\cite{armenio}},{\cite{gubser}})
 it was suggested, in the holographic scheme, a 
generalization of this theorem to four dimensional quantum field
theories admiting a five dimensional gravitational description. The idea
is of course to relate the function $c$ with some geometrical property
of the string metric and to use the sigma model beta function equations
to derive the desired behavior for $c$. The most natural candidate for $c$
is the expansion parameter ${\theta}$ for a congruence of
null geodesics. In terms of the gauge beta function we get for the universal
metric the following behavior of ${\theta}$
\be\label{theta}
{\mu}\frac{d{\theta}}{d{\mu}} = F_{U}({\mu})\frac{1}{g^{3}}{\beta}(g)
\ee
with $F_{U}({\mu})$ a universal function independent of beta
given by
\be
F_{U}({\mu}) = \frac{3}{2}(ln{\mu})^\frac{5}{2}
\ee
As it is clear from ({\ref{theta}}) the sign of variation
of ${\theta}$ depends on the sign of the beta function.
\par

To conclude just a comment on possible string corrections
to the sigma model beta function equations. We know that tadpoles beyond
tree level will in general modify the form of the background string metric
by a Fishler-Susskind 
{\cite{susskind}} type of mechanism. The difference between it and
the RGA  stems
precisely from the interpretation of the
renormalization group scale as a spacetime coordinate. 
This in particular implies
that the effect of tadpoles is absorbed in the dilaton dependence
on this coordinate; but we have shown
that this dependence can be always integrated in the universal metric.
The problem of tachyon stability of the string
metric worked out in this letter is analyzed in a different paper
see ref {\cite{alvarezgomez2}}.

%%%%%%%%%%%%%%%%%%%%%%%%%%%%%%%%%%%%%%%%%%%%%%%%%%%%%%%%%%%%%%%%%%%%%%
\section*{Acknowledgments}
%%%%%%%%%%%%%%%%%%%%%%%%%%%%%%%%%%%%%%%%%%%%%%%%%%%%%%%%%%%%%%%%%%%%%
This work ~~has been partially supported by the
European Union TMR program FMRX-CT96-0012 {\sl Integrability,
  Non-perturbative Effects, and Symmetry in Quantum Field Theory} and
by the Spanish grant AEN96-1655.  The work of E.A.~has also been
supported by the European Union TMR program ERBFMRX-CT96-0090 {\sl 
Beyond the Standard model} 
 and  the Spanish grant  AEN96-1664.

%%%%%%%%%%%%%%%%%%%%%%%%%%%%%%%%%%%%%%%%%%%%%%%%%%%%%%%%%%%%%%%%%%%%%%

%%%%%%%%%%%%%%%%%%%%%%%%%%%%%%%%%%%%%%%%%%%%%%%%%%%%%%%%%%%%%%%%%%%%%%

\end{document}